\documentstyle[aps,prb,preprint]{revtex}
\tighten

\begin{document}

\draft

\title{
Josephson current through a Luttinger liquid }
\author{
Rosario Fazio$^{(1)}$, F. W. J. Hekking$^{(2)}$,
and A. A. Odintsov$^{(2,3)}$
}
\address{
$^{(1)}$Istituto di Fisica, Universit\`a di Catania, viale A. Doria 6,
95129 Catania, Italy\\
$^{(2)}$Institut f\"ur Theoretische Festk\"orperphysik, Universit\"at
Karlsruhe, Postfach 6980, 76128 Karlsruhe, FRG\\
$^{(3)}$Present address:
Electrotechnical Laboratory, 1-1-4 Umezono, Tsukuba-shi,
Ibaraki 305, Japan\\
Permanent address:
Nuclear Physics Institute, Moscow
State University, Moscow 119899 GSP, Russia}

\date{\today}
\maketitle

\begin{abstract}
We study the Josephson effect through a one-dimensional system of
interacting electrons, connected to two superconductors by tunnel
junctions. The interactions are treated in the framework of the one-channel
Luttinger model.
At zero temperature, the Josephson critical current is found to decay
algebraically
with increasing distance between the junctions.
The exponent is proportional to the strength of the Coulomb interaction.
If the Luttinger liquid has a finite size, the Josephson current depends on
the total number of electrons  modulo 4.
These parity effects are studied for the ring,
coupled capacitively to a gate-voltage and threaded by a magnetic flux.
The Josephson current changes continuously as a function of the gate
voltage and {\em stepwise} as a function of the magnetic flux.
The electron-electron interaction
introduces {\em qualitatively} new features compared to the
non-interacting case.
\end{abstract}


Recent technological developments in the fabrication of semiconductor-
superconductor (S-Sc) interfaces made the observation of supercurrent in
S-Sc-S junctions possible~\cite{Nitta}. This
progress was accompanied by
a number of appealing theoretical predictions. The Josephson current
through a narrow constriction in a high-mobility two-dimensional
non-interacting electron gas (2DEG) should be quantized~\cite{Beenakker}.
Interference effects of two electrons entering the
normal region from a superconductor
are important, especially in mesoscopic samples~\cite{vanWees}.
This interference influences the Josephson current
through the normal region.

It is well-known that electron-electron interactions
affect the transport properties of mesoscopic devices as well.
The interaction manfests  itself in the single charge effects~\cite{Averin}.
In particular, the Josephson curent through SSS and SNS systems can be
modulated by a gate-voltage~\cite{Matveev}.
This has been described in terms of a phenomenological capacitance model.
However, in low-dimensional semiconductor nanostructures, with a low
electron density, electron-electron interactions should be treated
microscopically. As a result, the transport properties of a (quasi-)
one-dimensional quantum wire show deviations from
Fermi-liquid behaviour~\cite{Kane1}.

In the present paper we analyze the Josephson current through a Luttinger
liquid (LL).
Specifically, we consider two geometries which can be realized
experimentally:
A long wire with tunnel contacts to two
superconductors at a distance $d$ (Fig.~\ref{system}a), and a ring with
circumference $L$ and tunnel contacts at a distance $L/2$
(Fig.~\ref{system}b). Various aspects of transport in
mesoscopic systems (parity effects and interference combined with
electron-electron interactions) and their interplay can be studied in this
system.
The aim of this paper is twofold.
First, we want to study the influence of
the Coulomb interaction
on the Josephson critical current.
Second, we would like to see how
the parity effects which are present in the
ring~\cite{Haldane,Loss,Byers,Fujimoto}
manifest themselves in the dependence of the Josephson current on
magnetic flux and gate-voltage.
The coupling of bulk superconductors to a one-dimensional (1D)
system of interacting electrons
was studied by Fisher~\cite{Fisher}. In contrast to our work,
a chiral LL of spin-polarized electrons was considered and
tunneling with spin flip was analyzed.

The systems under consideration (Fig.~\ref{system})
 can be described by the Hamiltonian $\hat{H}
=
\hat{H}_{S1} + \hat{H}_{S2} + \hat{H}_L + \hat{H}_T . $
Here, $\hat{H}_{S1}$, $\hat{H}_{S2}$ are the BCS-Hamiltonians for the bulk
superconductors with gap $\Delta _{1}$, $\Delta _{2}$ respectively, kept at
a phase difference $\chi = \chi _1 -\chi _2$. The (1D)
electron system is described by the Hamiltonian~\cite{Kane2}
\begin{equation}
\hat{H}_L
=
\hbar
\int \frac{dx}{\pi }
\sum _{j}
v_j
\left[
\frac{g_j}{2} (\nabla \phi _j )^2 +
\frac{2}{g_j} (\nabla \theta _j)^2
\right] .
\label{lutham}
\end{equation}
It is written as a sum of the contributions from the spin ($j = \sigma $) and
charge ($j = \rho $) degrees of freedom.
We use the standard notation \cite{Kane2} for the parameters $g_j$ of the
interaction strengths and the velocities  $v_j$ of spin and charge
excitations.
We also introduce the bosonic fields
$\phi _{s}= \phi _{\rho} +s \phi _{\sigma}$ and
$\theta _{s}= \theta _{\rho} +s \theta_{\sigma}$ for spin up ($s=1$) and
down ($s=-1$) fermions. These fields obey the commutation relation $[\phi
_s (x), \theta _{s'} (x')] =
(i \pi /2) \mbox{ sign} (x'-x) \delta _{s,s'}$. The tunneling is assumed to
occur through two
tunnel junctions at the points $x=0$ and $x=d$,
\begin{equation}
\hat{H}_T
=
\sum _{s}
T _1 \hat{\Psi} ^{\dagger}_{S1,s} (x=0) \hat{\Psi}_{L,s}(x=0) +
T _2 \hat{\Psi} ^{\dagger}_{S2,s} (x=d) \hat{\Psi}_{L,s}(x=d) +
\mbox{(h.c.)}.
\end{equation}
The tunnel matrix elements
$T _{1,2}$ can be related to
the tunnel conductances $G_{1,2}$ of the junctions: $G_i = (4\pi e^2 /\hbar)
N_L(0) N_i (0) T_i^2 $,
where $N_L(0)= 1/\pi \hbar v_F$, $N_i$ is the normal density of states of a
superconductor $(i=1,2)$, and $v_F$ is the Fermi velocity.

The fermionic field operators $\hat{\Psi}$ can be expressed in terms of spin
and charge degrees of freedom ~\cite{Haldane}:
\begin{equation}
\hat{\Psi} ^{\dagger}_{L,s}(x,\tau)
=
\sqrt{\rho _{0,s}}
\sum \limits _{m, odd}
\exp \left\{i m k_F x \right\}
\exp \left\{i m \theta _{s} \right\} \exp \left\{ i \phi _s \right\},
\label{fieldoperator}
\end{equation}
where $k_F$ is the Fermi-wave vector and $\rho _{0,s} \equiv N_s/L =
k_F/(2\pi)$ is the average electron density for one spin direction. If the
LL is confined to a ring~\cite{Loss}, threaded by an Aharonov-
Bohm flux $\Phi$, the twisted boundary condition $\hat{\Psi}
^{\dagger}_{L,s}(x+L) = \exp{ \{- 2\pi i \Phi /\Phi _0\}} \hat{\Psi}
^{\dagger}_{L,s}(x)$ can be imposed~\cite{Byers}. Here, $\Phi _0$ is the
normal flux quantum $h/e$. The
fields $\theta $ and $\phi $ are decomposed in terms of
bosonic fields and topological excitations~\cite{Haldane} in the following
fashion:
\begin{eqnarray}
\theta _j(x)
& = &
\bar{\theta }_j(x) + \theta _j^0 + \pi M_j(x/2L) , \nonumber \\ \phi _j(x)
& = &
\bar{\phi } _j(x) + \phi _j^0 + \pi(J_j - 4 \delta_{j,\rho } \Phi/\Phi
_0)(x/2L) .
\label{substitution}
\end{eqnarray}
Here, $\bar{\theta }_j$ and $\bar{\phi }_j$ are the non-zero modes
\begin{eqnarray}
\bar{\theta} _j(x)
&=&
\frac{i}{2} \sqrt{\frac{g_j}{2}} \sum _{q \ne 0} \left|
\frac{\pi}{qL}
\right|^{1/2}
\mbox{sign}(q) e^{iqx}
(\hat{b}^{\dagger}_{j,q} +
\hat{b}_{j,-q}) , \nonumber \\
\bar{\phi} _j(x)
&=&
\frac{i}{2} \sqrt{\frac{2}{g_j}} \sum _{q \ne 0} \left|
\frac{\pi}{qL}
\right|^{1/2}
e^{iqx}
(\hat{b}^{\dagger}_{j,q} -
\hat{b}_{j,-q}) ,
\label{fields}
\end{eqnarray}
where $\hat{b}_{j,q}, \hat{b}^{\dagger}_{j,q}$ are Bose operators. $M_j$ and
$J_j$ denote the topological excitations. They are related to the usual
topological excitations for fermions with spin $s$: $M_s = (1/2)[M_{\rho} +
sM_{\sigma}]$ and $J_s = (1/2)[J_{\rho} + sJ_{\sigma}]$.
Using the topological constraints for $M_s$ and $J_s$ given in
Ref.~\cite{Haldane}, we find
constraints for $M_j$ and $J_j$:
(i) The topological numbers $M_j$ and $J_j$ are either {\sl simultaneously}
even or {\sl simultaneously} odd;
(ii) when $N_s$ is {\sl odd}
the sum
$M_{\rho} \pm M_{\sigma} + J_{\rho} \pm J_{\sigma}$ takes values $..., -4,
0, 4, ...$,
when $N_s$ is {\sl even}  the sum
$M_{\rho} \pm M_{\sigma} + J_{\rho} \pm J_{\sigma}$ takes values $...,-6,
-2, 2, 6, ...$.
Here, the number of electrons $N_s$
($N_{s=1} = N_{s=-1}$) determines the linearization point $k_F \equiv \pi
N_s/L$ of the original electron spectrum \cite{Loss}.

The Hamiltonian is
decoupled in the non-zero modes and the topological excitations:
\begin{equation}
\hat{H}_L
=
\hbar
\sum _{j=\rho ,\sigma}
\left\{
\sum _{q\ne 0}
v_j |q| \hat{b}^{\dagger}_{q,j} \hat{b}_{q,j} +
\frac{\pi v_j}{4 L}
\left[
\frac{g_j}{2}
(J_j - 4 \delta_{j,\rho } f_{\Phi})^2
+
\frac{2}{g_j} (M_j - 4 \delta_{j,\rho } f_{\mu})^2 \right]
\right\} .
\label{luttop}
\end{equation}
We introduced the flux frustration $f_{\Phi} \equiv \Phi/\Phi _0$,
as well as the parameter
$f_{\mu} = (g_{\rho}L/4\pi v_{\rho}) \Delta \mu$,
which is related to the difference $\Delta \mu$ between the electro-
chemical potential $\mu$ of the superconducting electrodes and the
Fermi energy $E_{F,0}$ of the ring.
For non-interacting electrons the latter is just the energy $(\hbar
k_F)^2/2M$ in the linearization point. Generally, the reference point $\Delta
\mu = 0$ is defined from the requirements that
for $\Phi = 0$
there are $2N_s$ electrons in the ground state ($N_1 = N_{-1}$) and
the energies to add and remove the electron(s) to/from the system are the
same. The electro-chemical potential $\mu$ can be controlled by a
gate-voltage.

The stationary Josephson effect can be obtained by evaluating the phase-
dependent part of the free energy $\cal{F}(\chi)$: The Josephson current is
given by
$
I_J
=
- (2e/\hbar) \partial \cal{F} / \partial \chi $.
We expand ${\cal F} = -(1 / \beta ) \ln Z$ where $Z = \mbox{Tr } \exp \{-
\beta \hat{H}\}$ in powers of $\hat{H}_T$;
the lowest order contribution arises in 4$^{th}$ order. It is represented by
the diagram in the inset of Fig.~\ref{jog}, which has a clear physical
meaning: the Josephson effect occurs through a process which transfers a
Cooper pair from superconductor $S2$, described by the anomalous Green's
function $F_{S2}(\tau _3, \tau_4)$ through the second
junction with an amplitude $(T_2^*) ^2$. Subsequently, the two electrons
propagate as a Cooperon through
the LL from contact 2 to 1.
This propagation is determined by the Cooperon propagator
$
\Pi (0,d;\tau _1, ...,\tau _4)
$.
Finally, both electrons tunnel through the first junction (amplitude $T_1^2$)
and enter the superconductor $S1$ as a Cooper pair, characterized by the
anomalous function $F_{S1}^{\dagger}(\tau _1, \tau_2)$.
If the distance $d$ between the junctions is much larger than the coherence
length in the superconductor, the characteristic energies $\hbar v_F/d$ of
the electrons propagating through the 1D system are much less than the
energies $\Delta$ of excitations in superconductors. At low temperatures
($k_BT \ll \Delta$), a generic process consists of fast tunneling of two
electrons from the superconductor into the 1D system ($|\tau_1-\tau_2|
\sim \hbar/\Delta$) and their slow propagation through the LL ($|\tau_1-
\tau_3| \sim d/v_F$).
The phase-dependent part of the free energy then simplifies to
\begin{equation}
{\cal F}(\chi)
=
- 2 \pi ^2 N_1 (0) N_2 (0)
\Re \mbox{e}[
T _1 ^2 (T_2^*)^2 e^{-i\chi}
\int _0 ^{\beta} d\tau \Pi(d; \tau)
] ,
\end{equation}
whith the Cooperon
\begin{equation}
\Pi(d,\tau)
=
\left\langle
\mbox{T}_\tau
\hat{\Psi}_{L,+}(0,0)
\hat{\Psi}_{L,-}(0,0)
\hat{\Psi} ^{\dagger}_{L,-} (d,\tau)
\hat{\Psi} ^{\dagger}_{L,+} (d,\tau)
\right\rangle ,
\label{cooperon}
\end{equation}
where the average is taken over the eigenstates of $\hat{H}_L$. The
evaluation of~(\ref{cooperon}) with the help of bosonized field operators
like~(\ref{fieldoperator}) is straightforward.

\noindent
{\it Wire geometry}.
For an infinitely long wire (Fig.~\ref{system}a), the topological excitations
play no role since their energies $\sim \pi v_j/L$ are vanishingly small. The
Cooperon propagator $\Pi (d,\tau)$ is given by
\begin{equation}
\Pi _w(d,\tau)
=
\rho _{0}^2
\sum _{n_1,n_2, odd}
e^{i(n_1+n_2)(k_Fd + \eta)}
\times
\left[
\frac{\alpha ^2}{d^2 + v_{\sigma}^2\tau ^2} \right]^{g_{\sigma} (n_1-
n_2)^2/16}
\left[
\frac{\alpha ^2}{d^2 + v_{\rho}^2\tau ^2} \right]^{1/g_{\rho} + g_{\rho}
(n_1+n_2)^2/16},
\label{cooperonwire}
\end{equation}
where
$e^{i\eta} =[(d+iv_{\rho}\tau)/(d-iv_{\rho}\tau)]^{1/2}$
, and $\alpha$ is a cut-off parameter~\cite{Luther} of the order $1/k_F$.
If $n_1 + n_2 \neq 0$, the rapid oscillations related to
$e^{i(n_1+n_2)(k_Fd)}$ make the Josephson term vanishingly small if the
tunnel junctions are large compared to the Fermi wavelength.
For $n_1 + n_2 = 0$ the leading contributions to the correlation function
correspond to $n_1 - n_2 = \pm 2$.
For the spin-independent electron-electron interaction we should fix
$g_{\sigma} = 2$ and $v_{\rho} = (2/g_{\rho})v_F$. For the Josephson
current at zero temperature we find
\begin{equation}
I_J
=
\frac{2 \pi e v_F}{d}
\frac{G_1 G_2}{(4e^2/\hbar)^2} F_w(g_{\rho},d) \sin (\chi) ,
\end{equation}
\begin{equation}
F_{w}
=
\left[\frac{1}{k_Fd}\right]^{2/g_{\rho} - 1}
\int \frac{dx}{\pi}
\frac{1}{\sqrt{1 + x^2}}
\left[\frac{1}{1+(2x/g_{\rho})^2}\right]^{1/g_{\rho}}
\end{equation}
($F_w = 1$ for noninteracting electrons).

\noindent
{\it Ring geometry}. For the ring  (Fig.~\ref{system}b) we get
\begin{equation}
I_J
=
\frac{2 \pi e v_F}{L}\frac{G_1 G_2}{(4e^2/\hbar)^2} \sum _{\epsilon = \pm
1}
\left\langle
F_{r}(g_{\rho},L,\epsilon, M_{\sigma}, J_{\rho})
\sin{(
\chi
+ \epsilon \pi M_{\sigma}/2 + \pi J_{\rho}/2)} \right\rangle _{J,M} ,
\label{josring}
\end{equation}
with
\begin{equation}
F_{r}
=
\left[
\frac{\pi}{k_FL}
\right]^{2/g_{\rho} - 1}
\int dx
\frac{1}{\cosh(x)}
\left[
\frac{1}{\cosh(2x/g_{\rho})}
\right]^{2/g_{\rho}}
\cosh
\left[
\left(\frac{2}{g_{\rho}}\right)^2
(M_\rho - 4f_{\mu}) x
+
\epsilon
J_\sigma x
\right] .
\label{fring}
\end{equation}
Here, $\langle ... \rangle _{J,M}$ means evaluation with respect to the
ground-state configuration of the topological excitations $J_j$, $M_j$ of
the LL subject to the topological constraints.

In Fig.~\ref{jog} the dependence of the critical current on
$g_{\rho }$ for the two geometries is shown.
The Josephson current is  suppressed by the repulsive
interaction, $I_J \propto d^{-2/g_{\rho}}$.
We estimate the magnitude of the current, using typical numbers from the
experiment of Mailly et al.~\cite{Mailly}, to be of the order of
several nA in the
non-interacting case.

We discuss now the flux and gate-voltage dependence in the ring geometry.
The Josephson current Eq.~(\ref{fring}) depends on the gate-voltage
via the parameter $f_{\mu}$. The flux does not enter expicitly into
Eq.~(\ref{fring}). However, the Josephson current depends on $f_{\Phi}$
(and on $f_{\mu}$) {\em implicitly} via the topological numbers
$(J_{\rho},J_{\sigma},M_{\rho},M_{\sigma})$. Therefore, one can expect
that the Josephson current changes stepwise as a function of the flux
(the jumps correspond to the change in the topological numbers) and shows
a continuous dependence (with jumps) on the gate-voltage
(see Figs.~\ref{joscritg2}, \ref{joscritg1.75}).

Without loss of generality we restrict the further consideration by
odd values of $N_s$~\cite{footnote4}.
We start from the non-interacting case $g_{\rho } = 2$.
The ground-state configurations
$(J_{\rho},J_{\sigma},M_{\rho},M_{\sigma})$
are displayed in Fig.~\ref{joscritg2}a.
Each state in the ring below the chemical potential $\mu$ of the
superconductors is occupied by two electrons with opposite spin. Hence,
the topological numbers $M_{\sigma}$ and $J_{\sigma}$
are always zero.
The increase of the flux  shifts down (up) the energy levels
of the electrons moving
in the positive (negative) direction along the ring.
Increasing $f_{\mu}$ correponds to a uniform shift of all energy levels
down with respect to the chemical potential $\mu$.
A {\em pair} of electrons tunnels into (out of) the ring each time when an
empty (filled) energy level crosses the chemical potential.
This leads to a change in the topological number $M_{\rho}$,
$\Delta M_{\rho} = 2 (-2)$. Simultaneously, the topological number
$J_{\rho}$ changes, $\Delta J_{\rho} = \pm 2$, since the  system acquires
orbital momentum. With the increase of the flux $\Phi$ by one flux quantum, a
pair of electrons enters the positive branch of the spectrum and (for a
different value of $\Phi$) another pair leaves the negative branch.
For each process $\Delta J_{\rho} = 2$. Accordingly, the Josephson critical
current
$I_{J,c} \equiv \max_{\chi} | I_J (\chi) |$
(Fig.~\ref{joscritg2})
changes stepwise with the flux making two jumps per period $\Phi_0$.
One of the jumps occurs at $f_{\Phi}
= -f_{\mu} + 1/2$ in  Fig.~\ref{joscritg2}.

For $f_{\mu} = 0$, the chemical potential of the superconductors $\mu$ is in
the middle of the gap between the last occupied and first unoccupied energy
levels ($E_{F,0} = \mu$). For this reason, the two processes mentioned above
occur at the same point $f_{\sigma} = 1/2$. One can say that a pair of
electrons jumps from the negative branch to positive branch
(e.g. the transition $(0,0,0,0) \rightarrow (4,0,0,0)$ in
Fig.~\ref{joscritg2}a)
and the number of electrons in the system is conserved.

The situation changes drastically for interacting electrons.
The interaction lifts
the degeneracy of the single particle states with different spin and the
system acquires additional stiffness with respect to the change in the
electron number (for the repulsive interactions $g_{\rho}<2$ the coefficient
$2/g_{\rho}$ of the term with the particle number
$M_{\rho}$ is larger than the coefficients of the  other topological terms in
Eq.~(\ref{luttop})). For this reason, the electrons tunnel {\em one by one}
into the ring and new ground-state configurations
(e.g. $(J_{\rho},J_{\sigma},M_{\rho},M_{\sigma}) =
(1,1,1,1)$ and $(3,-1,1 1)$) with an {\em odd} number of electrons arise
(see Fig.~\ref{joscritg1.75}a). In particular, the parity of the
electron number on the ring can be changed not only by changing
the gate-voltage, but also by changing the flux.
Depending on the value of the gate-voltage the increase of the flux by one
flux quantum is accompanied by
the tunneling of
two, one or zero electrons into the ring
and the same number of tunneling events out of it.
Accordingly, the Josephson current (Fig.~\ref{joscritg1.75}b) shows four,
two or zero jumps per period $\Phi_0$.

The size of the intervals of the gate-voltage, where the electron number
(characterized by $M_{\rho}$) does not depend on the flux, increases with
the increase of the repulsive interaction.
In the limit of strong interaction the number of particles is fixed for any
value of the gate-voltage except for small intervals near the points $f_{\mu} =
1/8 + n/4$ with integer $n$ \cite{footnote5}.

We have also found that
for some ground-state configurations, the Josephson current changes sign
and the ring acts as a $\pi $-junction.
This happens, for example, in the regions with $J_{\rho} = 2 \mbox{ mod}(4)$
both in the interacting and in the non-interacting cases
(e.g. in the region (2,0,2,0) in
Figs.~\ref{joscritg2}a, \ref{joscritg1.75}a).
Such behaviour
can be tested through the interference pattern of a SQUID, consisting of the
ring and a standard Josephson junction. A more detailed analysis of all these
features will be presented elsewhere~\cite{fazio}.

In this letter we studied the Josephson effect through a 1D
system of interacting electrons: a quantum wire and a ring.
The Josephson current was found to be supressed by
the Coulomb interaction.
The interaction qualitatively modifies the parity effects in the ring
and effects the dependences of the critical current on the flux and
gate-voltage. An anomalous sign of the Josephson current was found in some
range of parameters.

\noindent
{\bf Acknowledgements}
We would like to thank L. Glazman, D. Khmel'nitskii,
A.I. Larkin,
D. Loss, and G.T. Zimanyi for useful discussions. The financial support of the
European Community (HCM-network CHRX-CT93-0136) and of the Deutsche
Forschungsgemeinschaft through SFB 195 is gratefully acknowledged.

\begin{figure}
\caption{The geometries discussed in the text: (a) one-dimensional wire
connected to two superconductors by tunnel junctions, separated by a
distance $d$. (b)
Ring with circumference $L$, threaded by a magnetic flux $\Phi$. It is
connected to two superconductors by tunnel junctions, separated by a
distance $L/2$. \label{system}}
\end{figure}

\begin{figure}
\caption{Critical current as a function of $g_{\rho }$ for the ring (solid
line, normalized to $I_{J,c}^{(0)} = (4\pi
ev_F/L)(G_1G_2/(4e^2/\hbar)^2)$) and the wire (dashed line, normalized to
$I_{J,c}^{(0)} = (\pi
ev_F/d)(G_1G_2/(4e^2/\hbar)^2)$). Inset: Lowest-order contribution to the
Josephson effect. \label{jog}}
\end{figure}

\begin{figure}
\caption{(a) Ground-state configurations for topological
quantum numbers ($J_{\rho}$, $J_{\sigma}$, $M_{\rho}$, $M_{\sigma}$) as a
function of gate-voltage ($f_{\mu}$) and magnetic flux ($f_{\Phi}$), for
$g_{\rho} = 2$. (b) Critical current at $T=0$ for $g_{\rho} =2$ as a
function of $f_{\Phi}$ and $f_{\mu}$.
\label{joscritg2}}
\end{figure}

\begin{figure}
\caption{The same as Fig.~3, but for $g_{\rho } =1.75$.
\label{joscritg1.75}}
\end{figure}

\begin{references}
\bibitem{Nitta}J. Nitta et al., Phys.
Rev. B {\bf 46}, 14286 (1992); A. Dimoulas, J.P. Heida, B.J. van Wees,
T.M. Klapwijk, W. v.d. Graaf, and G. Borghs (unpublished)
\bibitem{Beenakker}C.W.J. Beenakker and H. van Houten, Phys. Rev. Lett. {\bf
66}, 3056 (1991); A. Furusaki et al., ibid. {\bf 67}, 132 (1991)
\bibitem{vanWees}B.J. van Wees et al., Phys. Rev. Lett. {\bf 69}, 510
(1992); F.W.J. Hekking and Yu.V. Nazarov, Phys. Rev. Lett. {\bf 71}, 1625
(1993)
\bibitem{Averin}D.V. Averin and K.K. Likharev, in ``{\sl Mesoscopic
Phenomena in Solids}'', edited by B.L. Altshuler, P.A. Lee, and R. Webb (North-
Holland, Amsterdam, 1991); G. Sch\"on and A.D. Zaikin, Phys. Rep. {\bf 198},
237 (1990)
\bibitem{Matveev}K.A. Matveev et al., Phys. Rev. Lett. {\bf 70}, 2940 (1993);
P. Joyez et al., ibid. {\bf 72}, 2458 (1994); R. Bauernschmitt et al.,
Phys. Rev.
B {\bf 49}, 4076 (1994)
\bibitem{Kane1}C.L. Kane and M.P.A. Fisher, Phys. Rev. Lett. {\bf 68}, 1220
(1992); K.A. Matveev and L.I. Glazman, ibid. {\bf 70}, 990 (1993)
\bibitem{Fisher}M.P.A. Fisher (unpublished)
\bibitem{Kane2}C.L. Kane and M.P.A. Fisher, Phys. Rev. B {\bf 46}, 15233
(1992)
\bibitem{Haldane}F.D.M. Haldane, Phys. Rev. Lett. {\bf 47}, 1840 (1981); J.
Phys. C {\bf 14}, 2585 (1981)
\bibitem{Loss}D. Loss, Phys. Rev. Lett. {\bf 69}, 343 (1992)
\bibitem{Byers} N. Byers and C.N. Yang, Phys. Rev. Lett. {\bf 7}, 46 (1961).
Note that the corresponding gauge transformation leads to an additional
phase factor for the tunnel matrix element $T_2$: $T_2 \to T_2 \exp{\{-i\pi
\Phi/\Phi_0\}}$ for the set-up depicted in Fig.~\ref{system}b.
\bibitem{Fujimoto}S. Fujimoto and N. Kawakami (unpublished)
\bibitem{Luther}A. Luther and I. Peschel, Phys. Rev. B {\bf 9}, 2911 (1974)
\bibitem{Mailly} D. Mailly et al. Phys. Rev. Lett. {\bf 70}, 2020 (1993)
\bibitem{footnote4} The case of even $N_s$ leads to a similar picture apart
from a relative shift along the $f_{\Phi}$ and $f_{\mu}$ axes.
\bibitem{footnote5}This regime resembles the situation in small metallic
tunnel junctions where the charging energy is much larger than the electron
level spacing.
\bibitem{fazio}R. Fazio, F.W.J. Hekking, and A.A. Odintsov (unpublished)
\end{references}
\end{document}